\documentclass{jltp}

\usepackage{graphicx} 

\title{Ballistic transport in classical and quantum integrable systems}

\author{Xenophon Zotos}

\address{Institut Romand de Recherche Num\'erique en Physique des
Mat\'eriaux,\\ PPH-Ecublens, CH-1015 Lausanne, Switzerland }

\runninghead{X. Zotos}{Ballistic transport}

\begin{document}

\maketitle

\begin{abstract}
In this essay, we first sketch the development of ideas on the 
extraordinary dynamics of integrable classical nonlinear 
and quantum many body Hamiltonians. In particular, we comment on the state 
of mathematical techniques available for analyzing their thermodynamic and 
dynamic properties. 

Then, we discuss the unconventional finite temperature transport    
of integrable systems using as example the classical Toda chain and 
the toy model of a quantum particle interacting with a fermionic bath in
one dimension; we focus on the singular long time asymptotic of 
current-current correlations, we introduce the notion of the Drude weight   
and we emphasize the role played by conservation laws in establishing 
the ballistic character of transport in these systems.

PACS numbers: 71.27.+a, 05.45, 72.10.-d
\end{abstract}

\section{INTRODUCTION}
The extraordinary stability of solitons upon collisions in integrable
nonlinear systems was first discovered in a numerical simulation 
of the Korteweg-de-Vries evolution equation, a system commonly studied 
in hydrodynamics and plasma physics\cite{zk}. This discovery was soon followed 
by the development of a beautiful mathematical theory - the Inverse Scattering 
Method (ISM) - that allows the analytical evaluation of the time evolution of 
an initial pulse configuration using linear operations\cite{ism}; this is the 
analogue of the Fourier Transform for linear systems.
These seminal works opened the new and still rapidly expanding field 
of nonlinear physics, with developments ranging from mathematical physics to 
applications\cite{os}.

Here, we should point out that there is a fundamental distinction between 
{\it integrable systems} (mostly one dimensional) characterized by the presence 
of ``mathematical solitons" and those with ``topological solitons". 
On the one hand, the stability of mathematical solitons is guaranteed by a 
subtle interplay between dispersion and nonlinearity; this interplay is 
expressed by the existence of a macroscopic number of 
{\it conservation laws} constraining the dynamical evolution. 
On the other hand, the stability of topological solitons 
is enforced by a topological constraint; of course there are examples, like 
the sine-Gordon field theory, which possess topological solitons but they 
are also integrable. In this work we will focus on the  
transport properties of integrable systems.

Due to the presence and stability upon scattering of nonlinear excitations, 
integrable systems are expected to show unconventional finite temperature 
transport properties, as ideal thermal, charge or spin 
conductivities, ballistic rather than diffusive transport.
Within the traditional framework of linear response theory,
the finite temperature dynamic correlations characterize the transport 
behavior and they are directly linked to experimental observations.
However, although integrable models are 
considered as exactly soluble, meaning that the initial 
value problem can be exactly analyzed using the ISM, rather little 
progress has been done in the evaluation of  
dynamic correlations which remains at best technically very involved.

In the quantum domain, 
parallel to developments on the analysis of classical integrable nonlinear 
systems, in the early 80's it was realized that the exact solution of a 
certain class of one dimensional quantum models by the Bethe ansatz (BA) method 
was equivalent to a quantum version of the ISM\cite{kor}.
In this class belong well known prototype systems as the Hubbard or spin-1/2
Heisenberg model, commonly used for the description of (quasi) one-dimensional
electronic or magnetic materials. 
The BA method provides the exact eigenfunctions and eigenvalues 
and by a certain procedure (and assumptions) the exact thermodynamic properties
and excitation spectrum. Similarly to their classical counterparts, the 
quantum systems possess a macroscopic number of conservation laws, 
characteristic of their integrability. 
It should therefore come as no surprise the proposition that quantum 
integrable systems should also exhibit unconventional transport\cite{czp}. 
The situation however is similar to the one of classical systems; 
although the exact eigenfunctions and eigenvalues are known, the 
calculation of finite temperature dynamic correlations is still out of 
reach for most of the models.

In the following, we discuss two simple examples, one classical and one 
quantum, in order to show the 
ideal conducting properties of integrable systems. 
Our strategy is to focus on the {\it long time} asymptotic 
value of the current correlations in order to demonstrate the ballistic 
transport instead of attempting to evaluate the full 
frequency dependence of the conductivity (or mobility). 
Furthermore, instead of linking the ideal conductivity 
to the dynamics of soliton excitations, not very transparent  
for quantum many body systems, we directly relate it to 
the conservation laws characterizing integrable systems. 

First, we present a study of the energy current - current correlations  
(related to the thermal conductivity) for the classical 
Toda chain. 
Second, we introduce the notion of the {\it Drude weight} as a criterion of 
ideal conductivity and evaluate it exactly using the Bethe ansatz 
method in the context of the mobility for a toy model describing a 
quantum particle interacting with a fermionic bath. 

\section{A classical system: the Toda chain}

The classical Toda lattice is a prototype model for studying   
the physics of nonlinear excitations\cite{toda}. It is one of the first 
models analyzed using the Inverse Scattering method\cite{flaschka}, 
the conservation laws characterizing this system\cite{henon} have been 
presented and it has even been invoked in attempts to describe nonlinear 
transport in DNA molecules\cite{dna}. 
As we mentioned above, 
although the initial value problem and the thermodynamic properties can 
be analytically studied, there is no clear picture on the finite 
temperature dynamic correlations\cite{ss}.

A physical quantity of interest in an anharmonic chain is the heat 
conductivity. In a generic case, it is expected that the 
energy current correlations decay to zero at long times and the decay is 
fast enough so that a transport coefficient can be defined. 
This behavior seems rather difficult to observe in several one dimensional 
systems\cite{livi,gr}, where the currents decay to zero but often too slowly,  
leaving the issue of diffusive transport controversial.

For an integrable model, ideal conducting behavior is expected 
with current correlations decaying to a finite value at long times.
To quantify the contribution of nonlinear excitations, different ingenious 
methods have been devised\cite{dna} (soliton counting procedures). 
Here, we will use the long time asymptotic value of current 
correlations as a measure of ideal transport, related in integrable systems 
to the existence of conservation laws. 

To establish this relation, we will use an inequality proposed 
by Mazur\cite{mazur}, linking the long time 
asymptotic of dynamic correlations functions to the presence of 
conservation laws: 
\begin{equation}
\lim_{T\rightarrow \infty} \frac{1}{T} \int_0^T \langle A(t)A\rangle 
dt \geq \sum_n
\frac{\langle A Q_n\rangle^2}{\langle Q_n^2\rangle} .
\label{mazur}
\end{equation}
Here $\langle \rangle$ denotes thermodynamic average,
the sum is over a set of conserved quantities ${Q_n}$, orthogonal
to each other $\langle Q_n Q_m\rangle=\langle Q_n^2\rangle\delta_{n,m}$ 
and we suppose that $\langle A\rangle=0$.

The classical Toda Hamiltonian for a chain of $L$ sites with periodic 
boundary conditions is given in reduced units by:
\begin{equation}
H=\sum_{l=1}^L \frac{p_l^2}{2}+e^{-q_l}
\label{toda}
\end{equation}
where $p_l$ is the momentum of particle $l$, $x_l$ its position and 
$q_l=x_{l+1}-x_l$. 

The energy current for a system of interacting particles is given 
by\cite{mcl}:
\begin{equation}
j^E=\sum_{l=1}^L p_l h_l+\frac{(p_{l+1}+p_l)}{2}q_l e^{-q_l} 
\label{je}
\end{equation}
where $h_l=\frac{p_l^2}{2}+\frac{1}{2}(e^{-q_l}+e^{-q_{l-1}})$.

We consider dynamic correlation functions in the fixed 
temperature-pressure thermodynamic ensemble:
\begin{equation}
\langle A(t)A\rangle=Z^{-1}\int \prod_{l=1}^L dp_l dq_l A(t)A e^{-\beta(H+PL)}
\label{ave}
\end{equation}
where $Z=\int \prod_{l=1}^L dp_l dq_l e^{-\beta(H+PL)}$, 
$L=\sum_{l=1}^L q_l$, $P$ is the pressure and $\beta$ the inverse 
of the temperature.

In this thermodynamic ensemble, equal time correlation functions can be 
calculated analytically. For instance the average distance is 
given by:
\begin{equation}
\langle q\rangle=\ln(\beta)-\Psi(\beta P)
\label{q}
\end{equation}
where $\Psi(z)$ is the digamma function.

The classical Toda lattice is characterized by a macroscopic number of 
conservation laws. The first few ones are:
\begin{eqnarray}
Q_1&=&\sum_{l=1}^L p_l\\
Q_2&=&\sum_{l=1}^L \frac{p_l^2}{2}+e^{-q_l}\\
Q_3&=&\sum_{l=1}^L \frac{p_l^3}{3}+(p_l+p_{l+1})e^{-q_l}\\
Q_4&=&\sum_{l=1}^L \frac{p_l^4}{4}+(p_l^2+p_lp_{l+1}+p_{l+1}^2)e^{-q_l}
+\frac{1}{2}e^{-2q_l}+e^{-q_l}e^{-q_{l+1}}\\
Q_5&=&\sum_{l=1}^L \frac{p_l^5}{5}+(p_l^3+p_l^2p_{l+1}+p_lp_{l+1}^2+p_{l+1}^3)
e^{-q_l}\\
&+&(p_l+p_{l+1})e^{-2q_l}+(p_l+2p_{l+1}+p_{l+2})e^{-q_l}e^{-q_{l+1}}...
\label{laws}
\end{eqnarray}
with $Q_1$ the total momentum, of course present in all 
translationally invariant systems, integrable or not, as also $Q_2$ the total 
energy. According to the standard Green-Kubo formulation of 
transport theory\cite{mcl} ``subtracted fluxes" should be used in the 
dynamic correlation functions determining the transport coefficients.
So in the case of energy transport we will study the decay of the 
``subtracted" energy current\cite{am}:
\begin{equation}
\tilde j^E=j^E-\frac{\langle Q_1 j^E\rangle}{\langle Q_1^2\rangle}Q_1
\label{jet}
\end{equation}
We see that the use of a subtracted flux is equivalent to removing the 
contribution of $Q_1$ in the long time asymptotic bound\cite{mazur} 
for $\langle j^E(t)j^E\rangle$.

We will now calculate a bound on the long time asymptotic value of 
$\langle j^E(t)j^E\rangle$ by the Mazur inequality eq.(\ref{mazur})
using the first $m$ conservation laws.  
We should note that $Q_3$ has a structure very similar to the energy 
current, so we expect a large contribution from this term; actually 
in some quantum models like the spin-1/2 Heisenberg chain or the t-J model,  
the energy current is identical to a conservation law, directly implying 
a nondecaying energy current and thus infinite thermal conductivity\cite{znp}.
Here, $Q_n's$ with $n=$even do not couple 
to $\tilde j^E$ so we will consider only $Q_n, n=3,5,7$. Higher 
$Q_n's$ can of course be included but the calculations become 
rather cumbersome.
Orthogonalizing the conserved quantities which appear in the right hand 
side  of eq.(\ref{mazur}) is equivalent to evaluating the expression:
\begin{equation}
C^m_{j^Ej^E}=\langle\tilde j^E|Q\rangle\langle Q|Q\rangle^{-1}
\langle Q|\tilde j^E\rangle
\label{ortho}
\end{equation}
where $\langle Q|Q\rangle$ is the $m\times m$ overlap matrix of $Q_n'n$ and 
$\langle Q|\tilde j^E\rangle$ the overlap vector of $\tilde j^E$ 
with the $Q_n's$.

\begin{figure}
\centerline{\includegraphics[height=3.0in]{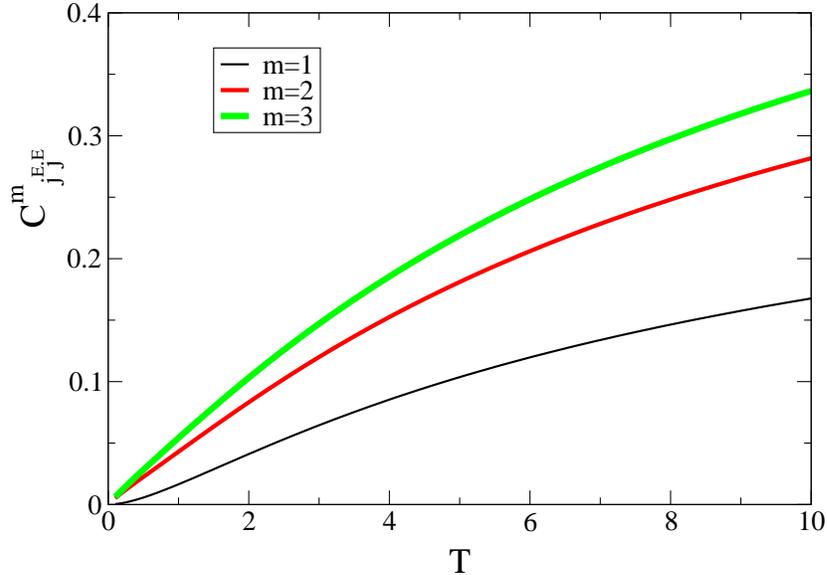}}
%
\caption{Lower bounds on the long time asymptotic value of energy current
correlations.}
\label{fig1}
\end{figure}

In Fig.1 we show the temperature dependence of 
$C^m_{j^Ej^E}/\langle\tilde {j^E}^2\rangle$ for $m=1~(n=3)$, $m=2~(n=3,5)$ and 
$m=3~(n=3,5,7)$. 
At low T the behavior is linear with slope $\frac{3}{35}$ for $m=2$ and 
$\frac{4}{63}$  for $m=3$. It is interesting that this value is comparable  
to the value for the density of solitons\cite{mb} 
$N_s/N=ln(2)/\pi^2 T$. So, we find that the long time asymptotic value of the 
subtracted energy current correlations is finite and most interestingly that 
it increases with temperature. This trend we can interpret as evidence for an 
increasing contribution of thermally excited nonlinear excitations on the 
ballistic transport.

The idea presented here provides, on the one hand a 
conceptual understanding of the role played by the conservation laws 
on the finite temperature dynamic correlations and on the other hand 
a simple analytical method for evaluating, or at least giving bounds on their 
long time asymptotic value. A similar analysis can be carried out for 
quantum integrable systems\cite{znp} although the complexity of the quantum 
conservation laws renders their wide use rather limited.

\section{A quantum system: the ``heavy particle" model}

The Drude weight $D$ (or charge stifness) was introduced as a 
criterion of an ideal conducting or insulating state at zero 
temperature\cite{kohn} and 
recently extended as a measure of ideal transport at finite 
temperatures\cite{czp}. Within linear response theory, 
it is essentially the prefactor of the low frequency 
reactive part of the conductivity,  
$\sigma''=2D/\omega|_{\omega \rightarrow 0}$, a finite $D$ 
implying a freely accelerating system. 
For a normal, diffusive system the Drude weight is zero at any finite 
temperature in the thermodynamic limit; according to the standard scenario 
the weight at zero frequency spreads to a ``Drude peak" with width 
proportional to the inverse of the collision time.
As we will see below, in integrable systems the Drude weight remains 
finite at all temperatures indicative of ballistic rather than diffusive 
transport.

The Drude weight can be conveniently evaluated\cite{kohn} as the thermal 
average of curvatures of energy levels $\epsilon_n$ of the system subject 
to a fictitious flux $\phi$,

\begin{equation}
D=\frac{1}{2L}\sum_n p_n \frac{\partial^2 \epsilon_n}{\partial \phi^2}|_
{\phi \rightarrow 0}
\label{drude}
\end{equation}
where $p_n$ are the Boltzmann weights and the sum is over all eigenstates of 
the system. It is also equal to the 
long time asymptotic value of the current-current correlations\cite{znp}, 

\begin{equation}
D=\frac{\beta}{2L}\langle j(t)j\rangle|_{t\rightarrow \infty}=
\frac{\beta}{2L} \sum_n p_n |\langle n|j|n\rangle|^2
\label{ltc}
\end{equation}
so useful bounds on $D$ can be obtained using the Mazur inequality following 
the same formulation as in the previous example.

For integrable quantum many body models it can evaluated exactly 
following recent developments in the Bethe ansatz technique, 
thus providing essential information on the transport 
properties of these systems without requiring the full calculation of the 
frequency dependence of the conductivity. 
This type of analysis, still under discussion as it is technically 
involved, has been carried out 
for several one dimensional integrable quantum models as the Hubbard chain, 
the spin 1/2 Heisenberg and the nonlinear-$\sigma$ model\cite{fk,f,xz}.
These calculations show that in most of the cases the Drude weight is 
finite at all temperatures implying ideal thermal, charge or spin 
conductivity. Recently, investigation of the finite temperature of these 
systems was also carried out by a semiclassical approach\cite{sachdev} and  
within the Luttinger liquid description\cite{gm,ra}.

Here, we will demonstrate the idea behind this type of Bethe ansatz analysis 
by evaluating the Drude weight related to the mobility of a quantum particle 
interacting with a bath of fermions in a one dimensional system\cite{mcguire}. 
A similar analysis was carried out for a particle moving on a lattice\cite{czp}
but the case discussed below is simpler and shows a qualitatively 
new behavior.  

We consider a particle with coordinate $y$ moving on a system of length $L$ 
with periodic boundary conditions and interacting with a set of N 
fermions described by the coordinates $x_j$ via a $\delta-$function 
interaction of strength $c$, 

\begin{equation}
H=-\sum_j\frac{\partial^2}{\partial x_j^2}
-\frac{\partial^2}{\partial y^2}+2c\sum_j\delta(x_j-y).
\label{hham}
\end{equation}

When the mass of the ``heavy particle" is equal to the mass of the fermions 
the model is integrable and so we expect a ballistic behavior of the 
mobility and therefore a finite Drude weight.
To evaluate $D$ using eq. (\ref{drude}), we 
consider the dependence of the energy levels 
on a flux $\phi$ acting only on the heavy particle.  
The momenta $k_j$ and the collective coordinate $\Lambda$ 
describing the Bethe ansatz wavefunctions are then given by the following 
standard equations obtained by applying periodic boundary conditions,

\begin{eqnarray}
Lk_j&=&2\pi I_j +\theta(k_j-\Lambda),~~~j=1,...,N+1,\\
\theta(p)&=&-2\tan^{-1}(2p/c),\\
L\sum_{j=1}^{N+1}k_j&=&2\pi\sum_{j=1}^{N+1}I_j+2\pi J+L \phi.
\label{pbc}
\end{eqnarray}
The eigenstates are characterized by the quantum numbers $(I_j,J)$ and their 
energy is given by:

\begin{equation}
E=\sum_{j=1}^{N+1} \epsilon(k_j)=\sum_{j=1}^{N+1} k_j^2.
\label{energy}
\end{equation}
These equations can be solved to order $1/L$ as we consider the effect 
of the one particle on the ensemble of fermions.
\begin{equation}
k_j=k_j^0+\frac{1}{L}\theta(k_j-\Lambda),~~~k_j^0=\frac{2\pi I_j}{L}
\label{k}
\end{equation}
Thus the total energy can be written as,
\begin{eqnarray}
&&E=\sum_j\epsilon(k_j^0)+\frac{2}{L} k_j^0 \theta(k_j^0-\Lambda),\\    
&&\frac{1}{L}\sum_j\theta(k_j^0-\Lambda)=\frac{2\pi J}{L}+\phi.
\label{o1}
\end{eqnarray}
Going to the continuum limit we obtain:
\begin{eqnarray}
&&E(\rho(k),\Lambda)=\frac{L}{2\pi}\int dk \rho(k) (k^2
+\frac{2}{L} k \theta(k-\Lambda)),\\
&&P+\phi=\frac{1}{2\pi}\int dk \rho(k) \theta(k-\Lambda), 
~~~P=\frac{2\pi J}{L}.
\label{cont}
\end{eqnarray}

Now we can define a correlation energy $\epsilon_c(\Lambda)$ 
assuming that the distribution of the fermion momenta is not affected 
by the presence of the extra particle and replacing the density 
$\rho(k)$ by the Fermi-Dirac distribution $f(k)$,

\begin{equation}
\epsilon_c(\Lambda)=\frac{1}{2\pi}\int dk f(k) 2k \theta(k-\Lambda).     
\label{ec}
\end{equation}
Using this formulation and the definition of the Drude weight eq.(\ref{drude})
we obtain: 

\begin{equation}
D=\frac{1}{2\pi Z_{\Lambda}}\int d\Lambda g(\Lambda) w(\Lambda) 
\frac{1}{2}\frac{\partial^2 \epsilon_c(\Lambda)}{\partial \phi^2}
\label{d}
\end{equation}
where, 
\begin{equation}
g(\Lambda)=\frac{\partial P}{\partial\Lambda}=\frac{1}{2\pi}
\int dk f(k) \frac{\partial \theta (k-\Lambda)}{\partial \Lambda},
\label{g}
\end{equation}

\begin{equation}
Z_{\Lambda}=\frac{1}{2\pi}\int d\Lambda g(\Lambda) w(\Lambda),~~~
w(\Lambda)=e^{-\beta\epsilon_c(\Lambda)}.
\label{z}
\end{equation}

\begin{figure}
\centerline{\includegraphics[height=3.0in]{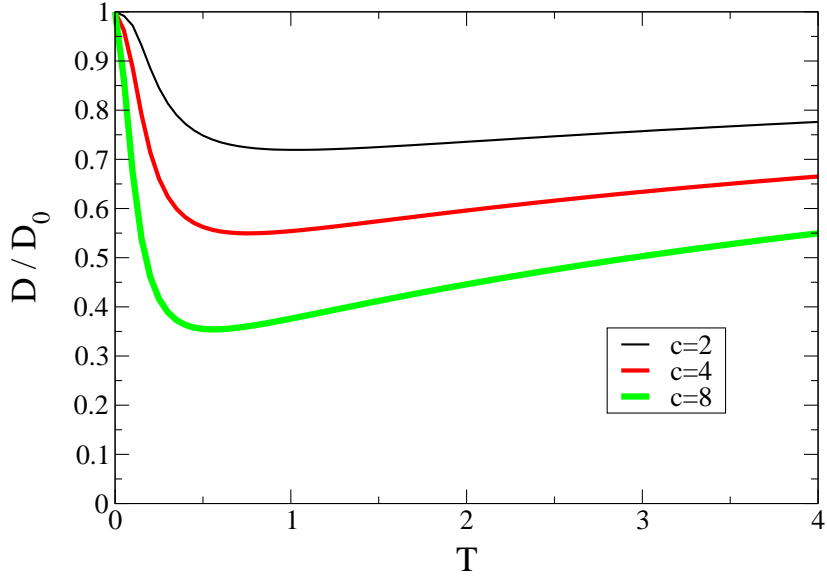}}
%
\caption{Drude weight of the ``heavy particle".}
\label{fig2}
\end{figure}

In Fig. 2 we show the normalized Drude weight $D/D_0$ for different values 
of the interaction $c$ as a function of temperature, with 
$D_0=D(T=0)=(\pi/2)(\tan^{-1}(2k_F/c)-(2k_F/c)/(1+(2k_F/c)^2))/
\tan^{-1}(2k_F/c)^2$ and $k_F=\pi n$.
The chemical potential is chosen so that we consider density $n=1$;
upon scaling $n\rightarrow nc, \beta\rightarrow \beta/ c^2$ 
the Drude weight $D$ remains the same.
We note that the behavior of $D$ is not monotonic, initially decreasing at low 
temperatures because of the interaction and then tending to the free 
particle value at high temperatures. This is in contrast to the Drude weight 
of systems on a lattice (tight binding models) where $D$ goes always to 
zero as $\beta$ at high temperatures.
The difference in behavior can be attributed to the bounded spectrum 
in lattice models in contrast to the unbounded one for continuous models.
Furthermore, numerical calculations on this model show that, 
the Drude weight vanishes at any finite temperature when the mass of the 
``heavy particle" is not equal to that of fermions, as expected for any 
normal system. 
In conclusion, the presented analysis demonstrates the basic features of  
the generic finite temperature ballistic transport behavior of integrable 
quantum many body systems.

\section*{ACKNOWLEDGMENTS}
It is my honor and pleasure to dedicate this paper to Professor Peter 
W\"olfle on the occasion of his 60th birthday.
This research was supported by the Swiss National Science Foundation, 
the University of Fribourg and Neuch\^atel.
I would also like to thank P. Prelov\v sek, H. Castella and F. Naef for 
our collaboration in the development of the ideas here presented.

\end{document}